\documentstyle[aps,prc,preprint,tighten]{revtex}

\begin{document}
\preprint{McGill/95-08}
\draft
\title{Measuring hadron properties at finite temperature}
\author{David Seibert\thanks{Electronic mail (internet):
seibert@hep.physics.mcgill.ca.}
and Charles Gale\thanks{Electronic mail (internet):
gale@hep.physics.mcgill.ca.}}
\address{Physics Department, McGill University, Montr\'eal, QC,
H3A 2T8, Canada}
\date{March 1995}
\maketitle
\begin{abstract}
We estimate the numbers and mass spectra of observed lepton and kaon
pairs produced from $\phi$ meson decays in the central rapidity region
of an Au+Au collision at lab energy 11.6 GeV/nucleon.  The following
effects are considered: possible mass shifts, thermal broadening due to
collisions with hadronic resonances, and superheating of the resonance
gas.  Changes in the dilepton mass spectrum may be seen, but changes in
the dikaon spectrum are too small to be detectable.
\end{abstract}

\section{Introduction}

There is currently a great deal of interest in finite-temperature
properties of hadronic resonances~\cite{ref1,ak1,ak2,ks,sk,hg,h}.
Recently, preliminary results for the $\phi$ meson mass spectrum in
Si+Au collisions at lab energy 11.6 GeV/nucleon were obtained at
Brookhaven's Alternating Gradient Synchrotron (AGS), by
reconstructing the $\phi$ mesons from final-state kaon pairs~\cite{bc}.
No change was observed from the vacuum width of the $\phi$, although
a possible small mass shift was observed for the most central events.
These negative preliminary results could dampen the enthusiasm of
other groups to study the $\phi$ peak in the dilepton channel.  We
show here that no observable change is expected for the $\phi$ peak
in the dikaon spectrum, but that effects may be visible in the
dilepton spectrum.  Thus, there is still good reason to study the
dilepton $\phi$ peak at the AGS, in spite of the fact that dikaon
results are negative.

\section{Equation of state}

The behavior of the hot matter is somewhat more complicated in
events at AGS energies than in ultra-relativistic events, due to the
large baryon densities and the lack of strangeness equilibration.
The equation of state used here is almost the same as those used in
Refs.~\cite{rcss,lthsr,kvv}.  We describe the system using the
temperature $T$, the baryon chemical potential $\mu_B$, and the
strangeness and antistrangeness chemical potentials, respectively
$\mu_S$ and $\mu_{\overline{S}}$, using the high energy conventions
$c = \hbar = k_B = 1$.

The matter has nearly isospin zero, so the up and down quark chemical
potentials, respectively $\mu_u$ and $\mu_d$, are $\mu_u=\mu_d=\mu_B/3$.
The $u$ and $d$ quarks are approximately in chemical equilibrium with
their antiquarks, so the $\overline{u}$ and $\overline{d}$ chemical
potentials are $\mu_{\overline{u}} = \mu_{\overline{d}} = -\mu_B/3$.
Finally, the net strangeness is also zero, since strong interactions
conserve strangeness and the time scales are too short to allow weak
interactions to be significant.  The $s$ and $\overline{s}$ chemical
potentials are respectively $\mu_s = \mu_S +\mu_B/3$ and
$\mu_{\overline{s}} = \mu_{\overline{S}} -\mu_B/3$, so that
$\mu_{\overline{S}} = \mu_S +2\mu_B/3$~\cite{sf}.
It is unlikely that $s$ and $\overline{s}$ are in chemical equilibrium
with each other, because $V R_s \ll n_s dV/dt$, where $V$ is the volume of
the hot matter, $R_s$ is the production rate of $s$ quarks per unit volume,
$n_s$ is the density of $s$ quarks, and $t$ is time.  However, we assume
for simplicity that the system is in chemical equilibrium, so that
$\mu_S=-\mu_B/3$; this assumption will be relaxed in later works.
The mean number of $s$ quarks per event is much greater than unity, so
the possible $s\overline{s}$ pair chemical potential \cite{sf} is omitted.

We model the QGP as a collection of free quarks and gluons, with a
constant (bag) energy density $B \simeq (220 \mbox{MeV})^4$.
The low-temperature phase is treated as a resonance gas (RG),
using all confirmed strongly-interacting particles
with known quantum numbers and without $c$ or $b$ content \cite{pdg}.
The chemical potential for resonance $i$ is
\begin{equation}
\mu_i = \left( \lambda_i^{(u)} +\lambda_i^{(d)} +\lambda_i^{(s)}
-\lambda_i^{(\overline{u})} -\lambda_i^{(\overline{d})}
-\lambda_i^{(\overline{s})} \right) \mu_B/3
+\lambda_i^{(s)} \mu_S
+\lambda_i^{(\overline{s})} \mu_{\overline{S}},
\end{equation}
where $\lambda_i^{(q)}$ is the number of constituent quarks of species
$q$ in resonance $i$, and the chemical potentials correspond to those
in the QGP.
We take an excluded volume, $v_{exc} = 4\pi R_{exc}^3/3$, for every
resonance in the RG phase \cite{rcss}.  Our justification for this is
that when the resonances overlap, the region where this occurs should be
in the QGP phase, so we do not allow overlap in the RG phase.

The transition temperature, $T_c$, is obtained by setting the pressures
equal in the two phases, with all chemical potentials fixed.  In Fig.~1,
we show the transition temperature as a function of $\mu_B$, for the
cases (i) $n_s=n_{\overline{s}}$, $\mu_{\overline{S}}=-\mu_S$
(strangeness equilibrium under the strong interactions), and (ii)
$\mu_{\overline{S}}=\mu_S \rightarrow -\infty$ (complete strangeness
suppression).  We give results for $R_{exc}=0.5$ and 1 fm in each case,
adjusting the bag constant to give $T_c=150$ MeV for $\mu_B=0$; for
the equilibrium case, we take $B^{1/4}=217$ and 220 MeV for
$R_{exc}=0.5$ and 1 fm respectively, while for the case of complete
strangeness suppression we take $B^{1/4}=207$ and 210 MeV.  Here (and for
the remainder of this paper) we evaluate all momentum integrals
numerically to one percent accuracy.

We fix $\mu_S$ for the strangeness equilibrium curves by requiring that
the QGP or RG be strangeness-neutral ($n_{\overline{s}}=n_s$).  At
fixed $\mu_B$, the transition from one strangeness-neutral phase to the
other phase (of arbitrary strangeness) occurs at almost exactly the same
temperature whether the initial phase is QGP or RG, so we only show RG
curves.  This result, first noted in Refs.~\cite{rcss,lthsr}, is somewhat
surprising, as the value of $\mu_S$ in the RG depends on the strange
hadron spectrum, so that the transition temperatures could easily be very
different for the two phases.

\section{Evolution of the hot matter}

We model the initial evolution following Ref.~\cite{s0}.  We approximate
the nucleon wavefunctions to be constant inside cylinders with radius
$r_N$ and length (parallel to the beam direction) $l_N$, and zero outside
these cylinders.  The radius is given by the nuclear radius, $r_N = 7$
fm for Au, while the length in the center of momentum (cm) frame is
$l_N \approx r_N/\gamma$.  Here $\gamma=(1-v^2)^{-1/2}$, where
$v \approx 1$ is the nuclear velocity in the cm frame; for fixed target
collisions at beam energy 11.6 GeV/nucleon, $\gamma=2.7$.  In principle,
$l_N$ is bounded from below by the minimum parton wavelength, of
order $m_{\pi}^{-1}=1.4$ fm, but for the collisions considered here
$l_N = 2.6$ fm so this lower bound is unimportant.

We then assume that the nucleon properties are unchanged during the
collision, and neglect collisions of secondary particles.  The rate of
production of any quantity during the collision is then proportional to
the overlap of the nucleon wavefunctions.  For example, if the cylinders
first touch at proper time $\tau=\sqrt{t^2-z^2}=0$, where $z$ is the
position along the beam axis, the baryon rapidity density at rapidity
$y=0$ is
\begin{equation}
dN_B/dy = k_B l_N^2 \times \left\{ \begin{array}{ll}
0, \quad                                  &\tau_* < 0,       \\
\tau_*^2, \quad                           &0 < \tau_* < 1/2, \\
\left[ 1 - 2(1-\tau_*)^2 \right]/2, \quad &1/2 < \tau_* < 1, \\
1/2, \quad                                &1 < \tau_*,
              \end{array} \right.
\end{equation}
where $k_B$ is an unspecified normalization constant, and
$\tau_*=\tau/l_N$.  The volume of hot matter per unit rapidity is
approximately $\pi r_N^2 \tau$, so the baryon density at $y=0$ is
\begin{equation}
n_B = \frac {dN_B/dy} {\pi r_N^2 l_N} \times \left\{ \begin{array}{ll}
0, \quad                                       &\tau_* < 0,       \\
2\tau_*, \quad                                 &0 < \tau_* < 1/2, \\
\left[ 1 - 2(1-\tau_*)^2 \right]/\tau_*, \quad &1/2 < \tau_* < 1, \\
1/\tau_*, \quad                                &1 < \tau_*,
              \end{array} \right.
\end{equation}
where $dN_B/dy$ is the observed baryon rapidity density at $y=0$.

For $\tau > l_N$, we evolve the hot matter hydrodynamically,
maintaining boost-invariance (and hence neglecting transverse expansion).
We assume strong superheating in the RG phase, so that the hot matter
never makes a transition to QGP, as the temperature never rises very
far above $T_c$.  The hot matter then follows the boost-invariant
hydrodynamic equations \cite{Bj}
\begin{eqnarray}
\frac {de} {d\tau} &=& \frac {-(e+P)} {\tau}, \label{ehyd1} \\
\frac {dn_B} {d\tau} &=& \frac {-n_B} {\tau}, \label{ehyd2} \\
\frac {d(n_s-n_{\overline{s}})} {d\tau} &=& 0, \label{ehyd3}
\end{eqnarray}
along with the equilibrium condition,
\begin{equation}
\mu_{\overline{S}} = -\mu_S. \label{eseqm}
\end{equation}
Here $e$, $P$, and $n_{\overline{s}}$ are respectively the energy density,
pressure, and $\overline{s}$ density.  We connect the hydrodynamic evolution
to the initial evolution by using the fact that the baryon and entropy
rapidity densities, respectively $dN_B/dy$ and $dS/dy$, are approximately
invariant after the initial collisions.

In Fig.~2, we show the trajectory followed by the hot matter at $y=0$
in a central Au+Au collision at 11.6 GeV/nucleon, using $R_f=0.5$ fm.
We use the preliminary results $dN_B/dy = 120$ and
$dN_{\pi^++\pi^-}/dy = 120$ \cite{fv} in the central rapidity region, from
which we estimate $dS/dy \sim 1700$ if the freezeout temperature
$T_f=100-120$ MeV.  At $\tau=l_N/2$ and $\tau=l_N$, $T=162$ MeV, and the
maximum temperature reached is 169 MeV, so the hot matter is only
superheated by $20-30$ MeV during the initial evolution.  For comparison,
we show also the trajectory obtained with the initial conditions used in
Ref.~\cite{kvv}, $e \simeq 2$ GeV-fm$^{-3}$ and $n_B \simeq 0.8$
fm$^{-3}$ for $\tau \leq l_N$, along with the critical temperature, $T_c$.
The maximum temperature reached with this second trajectory is 200 MeV,
significantly higher than for our trajectory.  However, our trajectory
should reflect the conditions of the thermalized matter, which is most
likely to remain at $y=0$, and thus will be accurately determined by
rapidity density measurements.

\section{Decay to observed lepton and kaon pairs}

There are a number of calculations of the $\phi$ mass and width at finite
$T$.  The range of predictions for the mass is very wide.  Haglin and
Gale~\cite{hg} find that the mass shift $\delta m(T)$ increases slowly
and monotonically with increasing $T$, with $\delta m(200~\mbox{MeV})
\approx 4$ MeV, while Asakawa and Ko~\cite{ak1,ak2} find that $\delta m$
decreases monotonically with $\delta m(190~\mbox{MeV}) \approx -170$ MeV.
The range of values for the width, $\Gamma(T)$, is much narrower.  Ko
and Seibert~\cite{ks} and Haglin~\cite{h} both find that the RG
contribution to $\Gamma(T)$ is approximately 30 MeV at $T=200$ MeV,
although it may be as low as 10 MeV if vertex form factors are
included~\cite{ks}.  The QGP contribution to the width depends on the
dynamics of the two-phase coexistence region, as it is proportional to
the fraction of matter in the QGP phase and inversely proportional to
the mean QGP droplet radius~\cite{sk}.  In the adiabatic limit (when the
nucleation rates for both phases are infinite), the QGP contribution to
$\Gamma(T)$ is infinite; however, the contribution to observable
quantities is probably small, just a few MeV in an Au+Au collision at
$\sqrt{s}=200$ GeV/nucleon and of order 10 MeV at $\sqrt{s}=10-20$
GeV/nucleon.

For the mass, we take two very different parametrizations, following
Refs.~\cite{hg} and \cite{ak1};
\begin{eqnarray}
m_{\phi}^{HG}(T) ~=~ 1020 \, + \, 4 (T/200~\mbox{MeV})^4~\mbox{MeV}, \\
m_{\phi}^{AK}(T) ~=~ 1020 \, - \, 200 (T/200~\mbox{MeV})^4~\mbox{MeV}.
\end{eqnarray}
We use these masses in the RG phase, and assume that there are no
recognizable $\phi$ mesons in the QGP.  As the corrections for finite
baryon density are uncertain, we neglect them at present.  For the
width, we use
\begin{equation}
\Gamma(T) ~=~ 4.43 \, + \, 25 (T/200~\mbox{MeV})^2~\mbox{MeV}.
\end{equation}

The dilepton signal from before freezeout is obtained by convoluting the
four-volume per unit rapidity, $\pi r^2 \tau d\tau$, the $\phi$ density
$n_{\phi}(T)$, the thermal phi mass distribution $p(m,T)$, and the decay
rate to dileptons, $\Gamma_{ll}$.  The post-freezeout contribution is
calculated in the same manner, except that we use the vacuum mass and
width, and assume that after freezeout the number of $\phi$ mesons
decreases exponentially in $\tau$ with the $T=0$ decay rate.
\begin{equation}
\frac {dN_{ll}} {dm^2 \, dy} ~=~ \pi r^2 \, \Gamma_{ll} \,
\left[ \frac {\tau_f \, n_{\phi}(T_f) \, p(m,0)} {\Gamma(0)} ~+~
\int_{\tau_0+l_N/2}^{\tau_f} d\tau \, \tau \, n_{\phi}(T) \, p(m,T) \right],
\end{equation}
We take $\Gamma_{ll}$ to be constant, evaluate $n_{\phi}$ numerically,
and assume a Breit-Wigner mass distribution,
\begin{equation}
p(m,T) = \frac {m_{\phi}(T) \, \Gamma(T) \, / \, \pi}
{\left[ m^2 \, - m_{\phi}^2(T) \right]^2
{}~+~ m_{\phi}^2(T) \, \Gamma^2(T)}.
\end{equation}
We neglect possible dileptons produced before the hot matter is
thermalized at proper time $\tau_0 \approx 1-2$ fm/$c$ after the point
of maximum overlap; this signal is
probably small, and we have no idea what the mass distribution is
for these particles.  The calculation is simplified by the fact that the
lepton mean free path is long, so that virtually all produced pairs
escape from the hot matter without interacting.

The kaon signal is calculated in almost the same manner, but with two
changes.  First, the decay rate to kaons, $\Gamma_{KK}$, depends strongly
on the $\phi$ mass.  We include this by using
\begin{equation}
\Gamma_{KK}(T) = \frac {m_{\phi}(0)^2 \,
\left[ m_{\phi}(T)^2 - 4 m_K^2 \right]^{3/2}}
{m_{\phi}(T)^2 \, \left[ m_{\phi}(0)^2 - 4 m_K^2 \right]^{3/2}}
\Gamma(0),
\end{equation}
where $m_K=495$ MeV is the vacuum kaon mass.  Second, we fold in the
probability for the kaons to escape the hot matter without interaction,
which is approximately
\begin{eqnarray}
{\cal P}_K(T) &=& \frac {2 \pi r_N \tau \times \tau_K(T) v_{\phi}(T)/8}
{\pi r_N^2 \tau}, \\
&=& \frac {\tau_K(T) v_{\phi}(T)} {4 r_N},
\end{eqnarray}
where $\tau_K$ is the mean time between kaon interactions, and
$v_{\phi}$ is the mean thermal velocity for $\phi$ mesons.  The numerator
in the first expression is the
product of the surface area of the cylinder and the mean distance from
the surface for a pair of emerging kaons (half that for a single kaon),
while the denominator is the total volume.  As the kaon velocity in the
$\phi$ rest frame is not too large, we assume that the kaon and $\phi$
velocities are the same; when the $\phi$ velocity is small enough that
this is not true, it is very unlikely that both produced kaons will
escape without interaction, so this should produce only a small
correction since our escape probability is proportional to $v_{\phi}$
and hence goes to zero in this case.
\begin{equation}
\frac {dN_{KK}} {dm^2 \, dy} ~=~ \pi r^2 \,
\left[ \tau_f \, n_{\phi}(T_f) \, p(m,0) ~+~
\int_{\tau_0+l_N/2}^{\tau_f} d\tau \, \tau \, n_{\phi}(T) \,
\Gamma_{KK}(T) \, {\cal P}_{KK}(T) \, p(m,T) \right].
\end{equation}
We take $v_{\phi}$ to be the (non-relativistic) rms thermal velocity,
\begin{equation}
v_{\phi}(T) ~=~ \left[ \frac {3 \, T} {m_{\phi}(T)} \right]^{1/2},
\end{equation}
and use the parametrization
\begin{equation}
\tau_K ~=~ 3 \, \left( \frac {150 \, \mbox{MeV}} {T} \right)^4 \,
\mbox{fm}/c,
\end{equation}
which agrees well with the results of Haglin and Pratt \cite{hp}.

We show predicted dilepton and dikaon mass spectra in Fig.~3, taking
$\tau_0=0$ and $T_f=100$ MeV.  The dikaon spectra differ significantly
only far from the $\phi$ peak, when the signal has dropped to less than
1\% of the peak value.  As there is large combinatoric background near
the peak, the small difference in the dikaon signal will almost
certainly be lost, while the difference in the dilepton signal is over
10\% and thus may be detectable.

Note the small secondary peak in the dilepton spectrum at $m \simeq 920$
MeV for $m_{\phi}^{AK}$ \cite{n1}.  This peak appears because the
hot matter spends $1-2$ fm/ with $T \simeq 160$ MeV, just as in the case
of a first-order phase transition \cite{ak1,ak2} (when the hot matter
remains at $T_c$ for a long time).  However, the peak here reflects the
(approximately constant) initial temperature, which results from the
spatial extent of the parton wavefunctions, and depends only weakly on
$T_c$.  In the case of a phase transition, the temperature of the hot
matter does not remain constant in the coexistence region even though
$T_c$ is the same for the two phases, because the entropy per baryon is
significantly higher in the QGP at fixed $T$ and $\mu_B$~\cite{hlr},
leading to smearing of any secondary $\phi$ meson peak created during a
phase transition at finite $\mu_B$.  The larger peak at $m \simeq 1000$
MeV is due to the fact that the system is assumed to freeze out
instantaneously, so that there is no hot matter with $0 < T < T_f$ and
thus no $\phi$ mesons are seen in the corresponding mass range; this
peak will soften and possibly disappear if more realistic dynamics are
used.

\section{Conclusions}

Although the total signal from dikaons is larger than that from
dileptons, the vast majority of the dikaons come from $\phi$ mesons
that decay after freezeout of the hot matter.  Thus, we see that, not
very surprisingly, it is difficult to measure thermal mass shifts by
observing strongly interacting particles.  Conversely, dileptons make a
good probe of thermal mass shifts, since they escape from the hot
matter throughout the collision.

There are two reasons for the lack of a strong signal from dikaons.
The first is that, once the $\phi$ mass drops appreciably, the decay
rate to dikaons vanishes.  However, this is not so easily corrected by
observing meson decay to lighter particles, such as pions.  Although the
decay rate then increases, the velocity of the decay products also
increases, so that it is greater than the mean thermal velocity of the
heavy meson.  The decay products then tend to emerge back-to-back in the
lab frame, so that the probability of both decay products escaping from
the hot matter without interacting is small.  It is also possible that
the kaon mass drops with increasing $T$~\cite{ls}, and that the $\phi$
mesons decay to these lighter kaons.  However, in this case we expect
that the thermal kaons will interact when they leave the hot matter, as
they acquire the vacuum kaon mass, and that this interaction will likely
remove the pairs from the $\phi$ peak.

As a result of the lack of sensitivity of the dikaon signal to thermal
$\phi$ meson properties, it is very important to study dilepton
production in ultra-relativistic heavy ion collisions, particularly
near vector meson peaks.  This is the only likely means for detecting
finite-temperature shifts in meson masses and widths.  As the study of
meson properties may provide valuable insight into the equation of
state of hot strongly-interacting matter \cite{sc,smf,ak1,ak2,ks,sk},
we would hope to see more experimental attention devoted to these
measurements in the future.

\acknowledgements

This work was supported in part by the Natural Sciences and Engineering
Research Council of Canada, and in part by the FCAR fund of the
Qu\'ebec government.  We thank J. Barrette, E. O'Brien, F. Videbaek, and
B. Cole for useful comments.

\newpage
\section*{Figure Captions}

\begin{description}
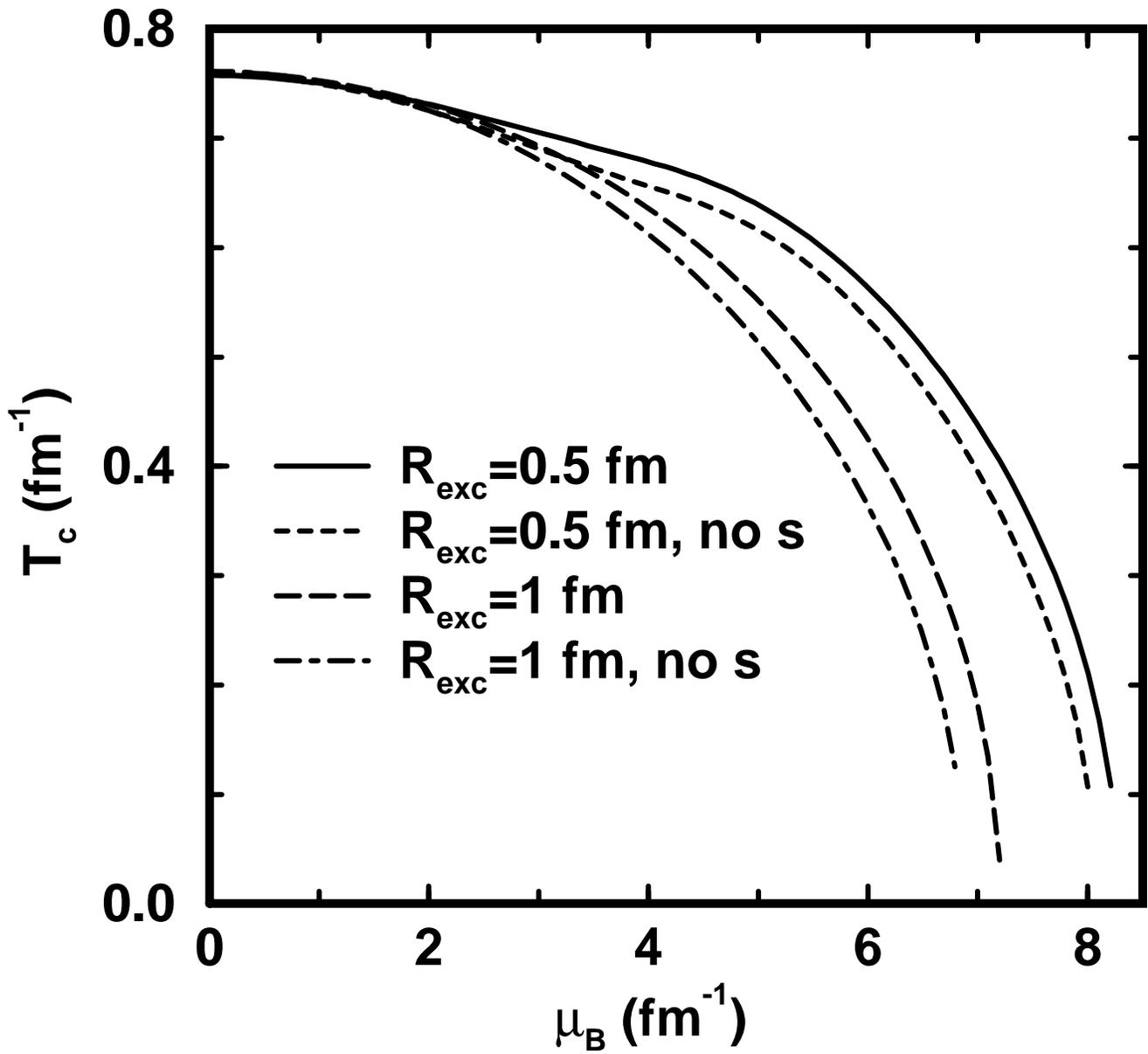
\item[Fig.~1:] $T_c$ vs.\ $\mu_B$ using $R_{exc}=0.5$ and 1 fm, for the
cases of strangeness equilibrium under the strong interactions and of
complete strangeness suppression (denoted ``no s'').
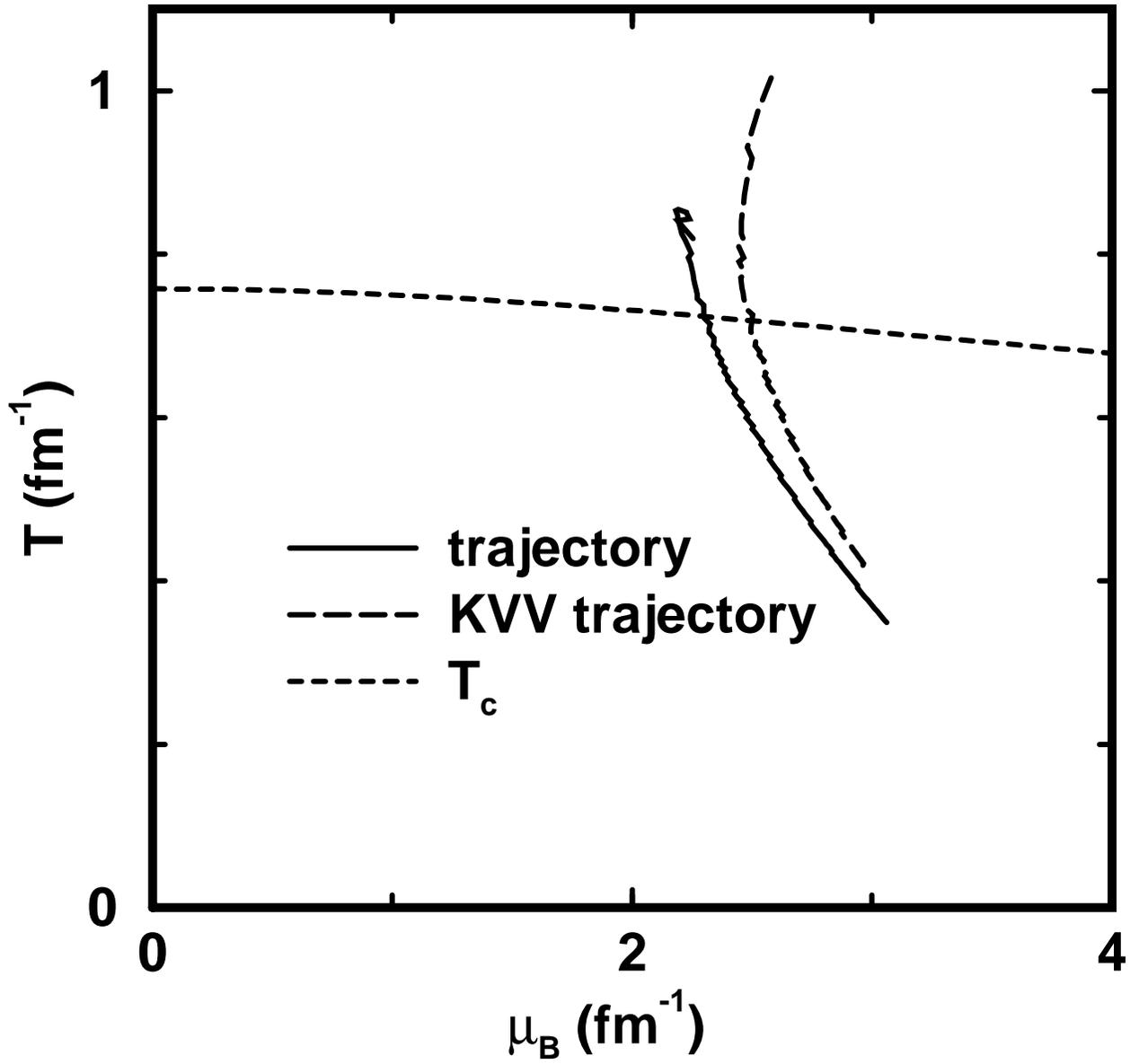
\item[Fig.~2:] Trajectories followed by the hot matter produced in a
central Au+Au collision at lab energy 11.6 GeV/nucleon.  The solid
trajectory follows the evolution described in the text, while the KVV
trajectory uses the initial conditions from Ref.~\cite{kvv} as the
starting point for the hydrodynamic evolution.  The matter is assumed
to remain in the RG phase, with $R_{exc}=0.5$ fm.  For comparison,
$T_c$ is shown for a chemically equilibrated resonance gas with
$n_s=n_{\overline{s}}$, taking $B=(217~\mbox{MeV})^4$.
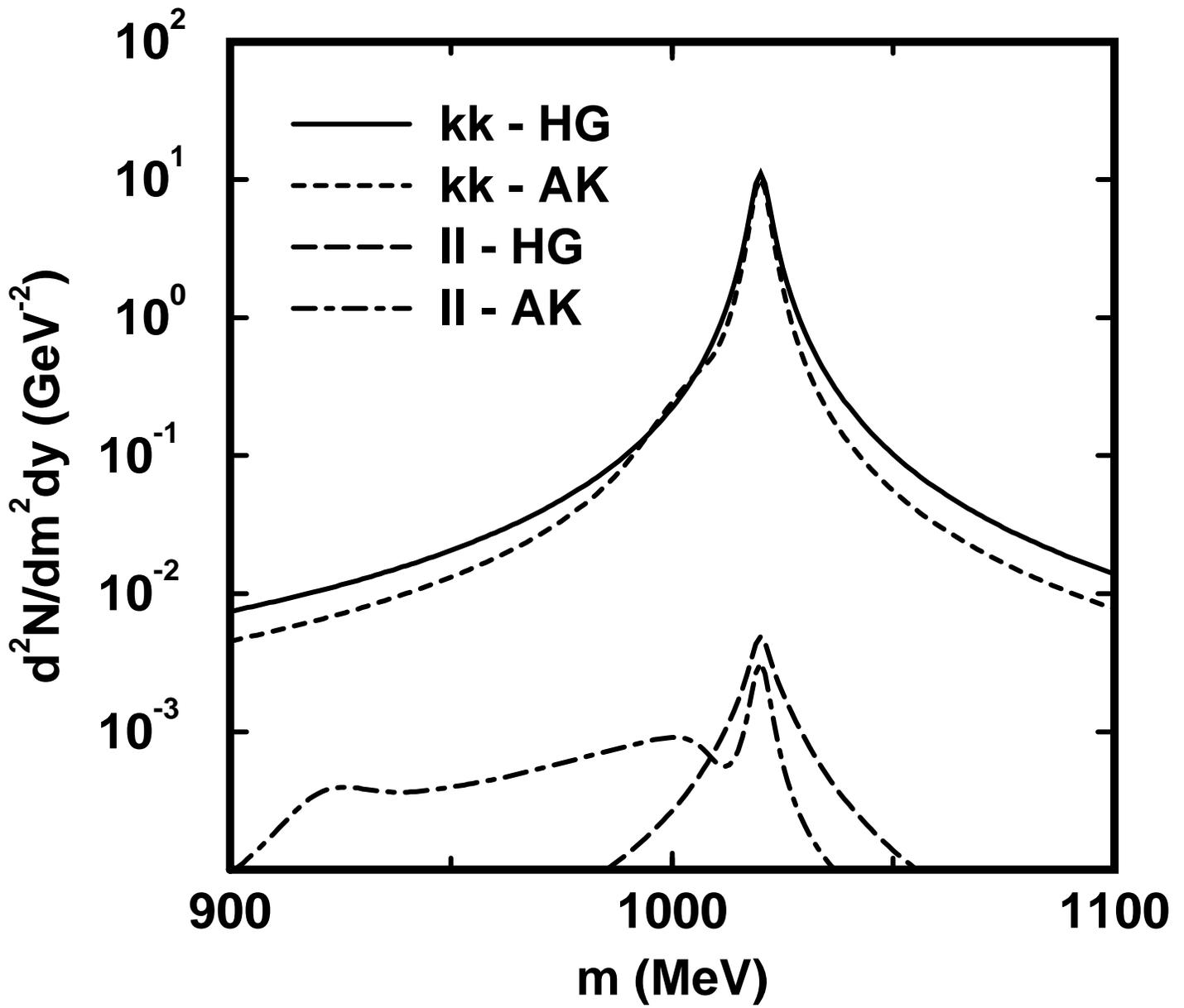
\item[Fig.~3:]  Predicted dikaon (kk) and dilepton (ll) spectra from a
central Au+Au collision at lab energy 11.6 GeV/nucleon, using
$m_{\phi}(T)$ from Refs.~\cite{hg} and \cite{ak1}, denoted by HG and AK
respectively.  Evolution of the hot matter proceeds as described in the
text, with $R_{exc}=0.5$ fm.
\end{description}

\vfill \eject


\begin{references}
\bibitem{n1} This peak is seen whenever $m_{\phi}(T_0)$ differs
significantly from $m_{\phi}(0)$, where $T_0$ is the initial
temperature (at $\tau=l_N$).
\end{references}
\end{document}